\documentclass[useAMS,usenatbib]{mnras}
\usepackage{amssymb}
\usepackage{amsfonts}
\usepackage{tabularx}
\usepackage{graphicx}
\usepackage{ulem}
\usepackage{array}
\usepackage{color}
\usepackage{booktabs}
\usepackage[fleqn]{amsmath}
\makeatletter
\newcommand*{\rom}[1]{\expandafter\@slowromancap\romannumeral #1@}
\makeatother
\numberwithin{equation}{section}
\def\d{{\rm d}}

\newcommand{\p}{\partial}

\newcommand{\case}{\textstyle\frac}

\newcommand{\ee}{\end{equation}}
\newcommand{\be}{\begin{equation}}

\begin{document}

\title[The Lynden-Bell bar formation mechanism]{The Lynden-Bell bar formation mechanism \\in simple and realistic galactic models}

\author[E.\,V.\,Polyachenko \& I.~G.~Shukhman]{
E.~V.~Polyachenko,$^{1}$\thanks{E-mail: epolyach@inasan.ru}
I.~G.~Shukhman,$^{2}$\thanks{E-mail: shukhman@iszf.irk.ru}
\\
$^{1}$Institute of Astronomy, Russian Academy of Sciences, 48 Pyatnitskya St., Moscow 119017, Russia\\
$^{2}$Institute of Solar-Terrestrial Physics, Russian Academy of Sciences, Siberian Branch, P.O. Box 291, Irkutsk 664033, Russia
}


\maketitle
\pagerange{\pageref{firstpage}--\pageref{lastpage}} \pubyear{2020}

\maketitle

\begin{abstract}

Using the canonical Hamilton-Jacobi approach we study the Lynden-Bell concept of bar formation based on the idea of orbital trapping parallel to the long or short axes of the oval potential distortion. The concept considered a single  parameter~-- a sign of the derivative of the precession rate over angular momentum, determining the orientation of the trapped orbits. We derived a perturbation Hamiltonian which includes two more parameters characterising the background disc and the perturbation, that are just as important as the earlier known one. This allows us to link the concept with the matrix approach in linear perturbation theory, the theory of weak bars, and explain some features of the nonlinear secular evolution observed in N-body simulations.

\end{abstract}


\section{Introduction}

A remarkable paper by \cite{1979MNRAS.187..101L} had influenced the bar formation theory in stellar discs and the radial-orbit instability theory in spherical clusters~\citep[e.g.][]{2015MNRAS.451..601P}. It considers a weak oval distortion of the potential (bar) rotating with pattern speed $\Omega_{\rm p}$. A substantial group of stars within the corotation radius obey a condition
\be
\left|\Omega-\Omega_{\rm p} - {\case{1}{2}}\,\varkappa \right| \ll \Omega
\label{eq:slow_c}
\ee
to which we will refer below as `the slowness condition'. Here $\Omega$ and $\varkappa$ denote the angular speed and the epicyclic frequency of radial oscillations. In the reference frame of the bar, the motion of these stars can be viewed as slow nodal precession of stellar orbits, as long as the fast motion of stars along the orbits can be averaged out.

Lynden-Bell suggested a very elegant qualitative way to describe the dynamics of these orbits, in particular, their ability to align parallel to the long/short axis of the potential thereby reinforcing/weakening the primordial oval perturbation. Recall that orbits in such a weakly non-axisymmetric system possess a specific integral of motion $J_f = L/2 + I$, while the angular momentum $L$ and radial action $I$ of the star is changing. The key insight was that if the precession rate decreases/grows with $L$ (at constant $J_f$), such orbits seek for stationary position perpendicular/parallel to the bar. The former orbits were declared as `normal', whereas the latter declared  `abnormal' since they occupy only a small fraction of the phase space in the centre of the disc. Mathematically, the `normal'/`abnormal' orbits have negative/positive derivative of the precession rate over $L$ at constant $J_f$. Given the importance of this derivative in stellar dynamics (of the precession rate, distribution functions, etc.), we began to call it `LB-derivative'~\citep{EP04, EP05}.

Matrix methods of linear perturbation theory for study instability in the disc and spherical stellar systems show that sign of the precession rate is an important parameter. For instance, the loss cone instability~\citep{1991SvAL...17..371P, 2005ApJ...625..143T, 2007MNRAS.379..573P, 2008MNRAS.386.1966P} is sensitive to the sign of the precession rate itself, not to the sign of its derivative. On the other hand, \citet{1985AJ.....90.1027M}, and then \citet{1991MNRAS.248..494S, 1991ApJ...368...66W, 1994ASSL..185.....P} used the Lynden-Bell idea to explain a mechanism of the radial-orbit instability (ROI) in spherical systems. This idea indeed can be justified in the case of extremely slow ROI, although generally, it is invalid \citep[see details in][]{2015MNRAS.451..601P}. This hints to the existence of other parameters in addition to the LB-derivative of the precession rate governing the orbital alignment.

It is also worth noting that the theory of weak bars suggests orbits' alignment parallel to the long axis of the potential from the corotation resonance (CR) inside up to the centre or the inner Lindblad resonance (ILR) \citep[][hereafter BT]{1976ApJ...209...53S, BT08}. In contrast, the Lynden-bell mechanism (within the region of applicability inside CR) predicts the orbital alignment parallel to the short axis of the potential everywhere excluding a small central region where the LB-derivative is positive.

The goal of this paper is to analyse the problem consistently using a standard rigorous technique of finding stationary points parenting families of trapped orbits. Section 2 describes the technique in a short form. Section 3 contains two analytic examples (the power-law angular speed and the isochrone potential) and results of N-body simulations of a realistic Milky Way model. Finally, in Section 4 we discuss and summarise the results.

\section{The Hamiltonian-Jacobi approach for stationary points}

In this section, we employ the standard formalism to find families of orbits trapped by the bar potential. To this end, we find stationary points of the Hamiltonian equations that mark closed elliptical orbits parenting families of trapped orbits. These closed orbits are analogues of circular orbits in axisymmetric potentials. The stellar motion is considered in the epicyclic approximation, and the bar pattern speed $\Omega_{\rm p}$ obeys the slowness condition (\ref{eq:slow_c}).

The Jacobi integral for axisymmetric potential $\Phi_0(r)$ in the rotating frame can be written as
\begin{multline}
  \label{eq:HJ0}
  H_0(L,I) ={\case{1}{2}}\,\Omega^2(R)\,R^2 +\Phi_0(R) -\Omega_{\rm p}\,L + \\ + \varkappa\,(R)\,I+ \beta(R)\, I^2 \,,
\end{multline}
where $R=R(L)$ is the guiding centre radius. In order to obtain linear corrections $\propto {\cal O}(I)$ for the angular speed $\Omega(R)$ and the epicyclic frequency $\varkappa(R)$, we retain a small post-epicyclic term $\beta I^2$. An explicit form of $\beta$ can be found, e.g., in \cite{1969ApJ...158..505S},   \cite{1975ApJ...201..566C}, \cite{1976ApJ...203...81M} and \cite{2014dyga.book.....B}:
\be
  \beta=\frac{1}{8R^2}\Bigl(3q-\frac{1}{3}\,q^2+\frac{1}{2}\,R\,\frac{dq}{dR}\Bigr),\ \
  q = \frac{ d\ln(\varkappa^2)}{d\ln R}\,.
\ee
From (\ref{eq:HJ0}) we obtain:
\begin{align}
\Omega_1(L,I) &\equiv \frac{\p H_0(L,I)}{\p I}=\varkappa+2\beta\,I + {\cal O}(I^2)\,, \\
\Omega_2(L,I) &\equiv \frac{\p H_0(L,I)}{\p L}=\Omega-\Omega_{\rm p} +\frac{d\varkappa}{dL}\,I + {\cal O}(I^2)\, .
\label{eq:Omega_1_2}
\end{align}
The orbit precession rate in the rotating frame is
\be
  \Omega_{\rm pr}(L,I) \equiv \Omega_2(L,I)-{\case{1}{2}}\,\Omega_1(L,I)\,.
 \label{eq:Omega_pr}
\ee

Let's $\delta\Phi$ be a weak oval distortion of the axisymmetric disc potential $\Phi_0(r)$ rotating with pattern speed $\Omega_{\rm p}$,
\be
  \delta\Phi(r,\varphi)=A(r)\,\cos(2\varphi)\,,\quad A<0 \,.
  \label{eq:dPh}
\ee
This form suggests that troughs of the potential and crests of the perturbed surface density are oriented along the horizontal axis $OX$. A full Hamiltonian $H_J$ is then equal to a sum of the Jacobi integral (\ref{eq:HJ0}) and the perturbed potential (\ref{eq:dPh}), $H_J=H_0+\delta\Phi$.

Following~\citet{EP04, EP05}, we perform transformation of action-angle variables
\be
   I \to J_f = I + \case 12 L\,,\quad w_2 \to \phi = w_2 - \case12 w_1\,
   \label{eq:sac}
\ee
to benefit from having slowly varying angle variable $\phi$ compared to $w_1$, provided that $\Omega_{\rm p}$ obeys (\ref{eq:slow_c}). Averaging the Jacobi integral (\ref{eq:HJ0}) over $w_1$ gives a new integral of motion $J_f$. Using the epicyclic approximation,
\be
r = R - \rho \cos w_1\,, \quad \varphi = \phi  + \case12 w_1 + \displaystyle\frac{2\Omega}{\varkappa} \frac{\rho}R \sin w_1\,,
\label{eq:epic}
\ee
one can have for the averaged bar potential:
\begin{multline}
  {V}(L,J_f,\phi) = \frac{1}{2\pi}\,\oint \delta\Phi\Bigl(r(L,J_f,w_1), \varphi(L,J_f,w_1,\phi)\Bigr)\,dw_1 = \\
   = B(L,J_f)\,\cos(2\phi)\,,
\end{multline}
where $\rho = (2I/\varkappa)^{1/2}$ is the epicyclic radius,
\be
 B(L,J_f)=-\frac{A(R)}{2}\,\left(\frac{\rho}{R}\right)\,
 \left(\frac{R}{A}\frac{dA}{dR}+\frac{4\,\Omega}{\varkappa}\right)\,.
 \label{eq:bcap}
\ee
From (\ref{eq:sac}) and (\ref{eq:epic}) we infer that orbit's apocentre is parallel to the long axis of the potential if angle variable $\phi = \case 12 \pi$ or $\case32\pi$, and to the short axis if $\phi = 0$ or $\pi$.

Omitting the terms depending on $J_f$ only, one can end up with the following expression for the Ha\-mil\-tonian averaged over the fast orbital motion:
\begin{align}
{\cal H}_J(I,\phi) &=-2\,{\cal Q}\,I+2\,{\cal P}\,I^2+{V}(L,J_f,\phi) \notag \\
&=-2\,{\cal Q}\,I+2\,{\cal P}\,I^2+b(L)\,I^{1/2}\,\cos(2\phi)\,.
\label{eq:cal_H_J}
\end{align}
The coefficients ${\cal Q}$ and ${\cal P}$ are the precession rate of the orbits in the rotating frame and the LB-derivative of the precession rate in the limit of small $I$:
\begin{align}
{\cal Q} &\equiv \Omega_{\rm pr} (L, 0) \,, \label{eq:calQ} \\
{\cal P} &\equiv \left. \frac{\d {\cal Q} }{\d L}  - \frac12 \frac{\d \varkappa}{\d L} + \frac12 \beta \right.\, . \label{eq:calP}
\end{align}
If ILRs are present, ${\cal Q}$ is positive between the first (inner) and the second (outer) ILRs. In the absence or outside ILRs, ${\cal Q}$ is negative.
Factor $b I^{1/2}$ in the last term of the Hamiltonian substitutes the amplitude of the averaged bar potential $B(L,J_f)$ (see \ref{eq:bcap}). The new parameter
\be
b(L) = -\frac{A(R)}{2}\,\left[ \frac{2}{\varkappa R^2}\right]^{1/2}\,
 \left(\frac{R}{A}\frac{dA}{dR}+\frac{4\,\Omega}{\varkappa}\right)
 \label{eq:b}
\ee
reflects the orbital {\it responsiveness} to the bar-like perturbation.

Note that in fact ${\cal Q}$, ${\cal P}$, and $b$ are functions of invariants, so no derivation over $I$ is needed. Within the adopted approximation, however, these invariants can be replaced by $L$ (or $R$). To justify this, one needs to consider a small perturbation of the angular momentum, $h \equiv L-L_0$, near the angular momentum $L_0$ of the circular orbit on a given radius. The scaling adopted in this paper is the following: $h, I, Q ={\cal O}(\varepsilon^{2/3})$ and $P={\cal O}(1)$, where $\varepsilon$ is a small parameter characterising the oval distortion, i.e. $A={\cal O}(\varepsilon)$. In doing so, we obtain ${\cal Q} = \Omega_{\rm pr} (J_f, 0)$ and ${\cal P}= {\cal P}(L_0)$. Changing $J_f$ and $L_0$ in the arguments of these functions to $L$ gives additional terms of the order ${\cal O}(\varepsilon^2)$ which are smaller than all terms retained in the Hamiltonian $({\cal O}(\varepsilon^{4/3})$). The detailed derivation can be found in \citet{2020AstL...46...12P}.

Similar technique based on the averaged Jacobi Hamiltonian near ILR for spiral perturbations using the post-epicyclic approximation including the terms up to $(I^{1/2})^4$  was elaborated in \cite{1975ApJ...201..566C}, but it differs in some details. Apart from the different form of perturbation, there are distinctions in the derivation of the averaged Hamiltonian. In particular, Contopoulos considered $L_0$ as the angular momentum of stars exactly on ILR, while in our case, $L_0$ is the angular momentum of any orbit obeying (\ref{eq:slow_c}); the ILR may be absent. Besides, two small parameters of the problem -- the amplitude of the spiral potential $A$ and the epicyclic parameter $I (\sim h)$, were considered as independent ones, while in our case they are related by the scaling given above. The latter allows us to obtain the final results much easier.

Stationary points are derived from the equations:
\be
\frac{\p {\cal H}_J}{\p \phi}=0\,,\quad \frac{\p {\cal H}_J}{\p I}=0\,,
\label{eq:cal_SP}
\ee
which yield
\be
 \sin(2\phi)=0\,,\quad {\cal Q}\,I^{1/2} -2\,{\cal P}\,I^{3/2}-\case 14\, b\,\cos(2\phi)=0\,.
\label{eq:stat_points1}
\ee
Finally, we obtain the next conditions for the radial actions:
\begin{align}
 f(I^{1/2}) &= \phantom{-}\case 14 b \quad (\textrm{short axis: } \phi=0,\ \pi)\,, \label{eq:short} \\
 f(I^{1/2}) &= -\case 14 b \quad (\textrm{long axis: } \phi=\pi/2,\  3\pi/2)\,, \label{eq:long}
\end{align}
where $f(z) = {\cal Q}\,z - 2\,{\cal P}\,z^3$.

The negative sign of $b$ essentially occurs at the end of the bar, i.e. in the vicinity of the corotation, see discussion below. Thus we shall mainly assume $b>0$; the opposite case will be treated separately.

\begin{figure}
\centering
  \centerline{\includegraphics[width=\linewidth]{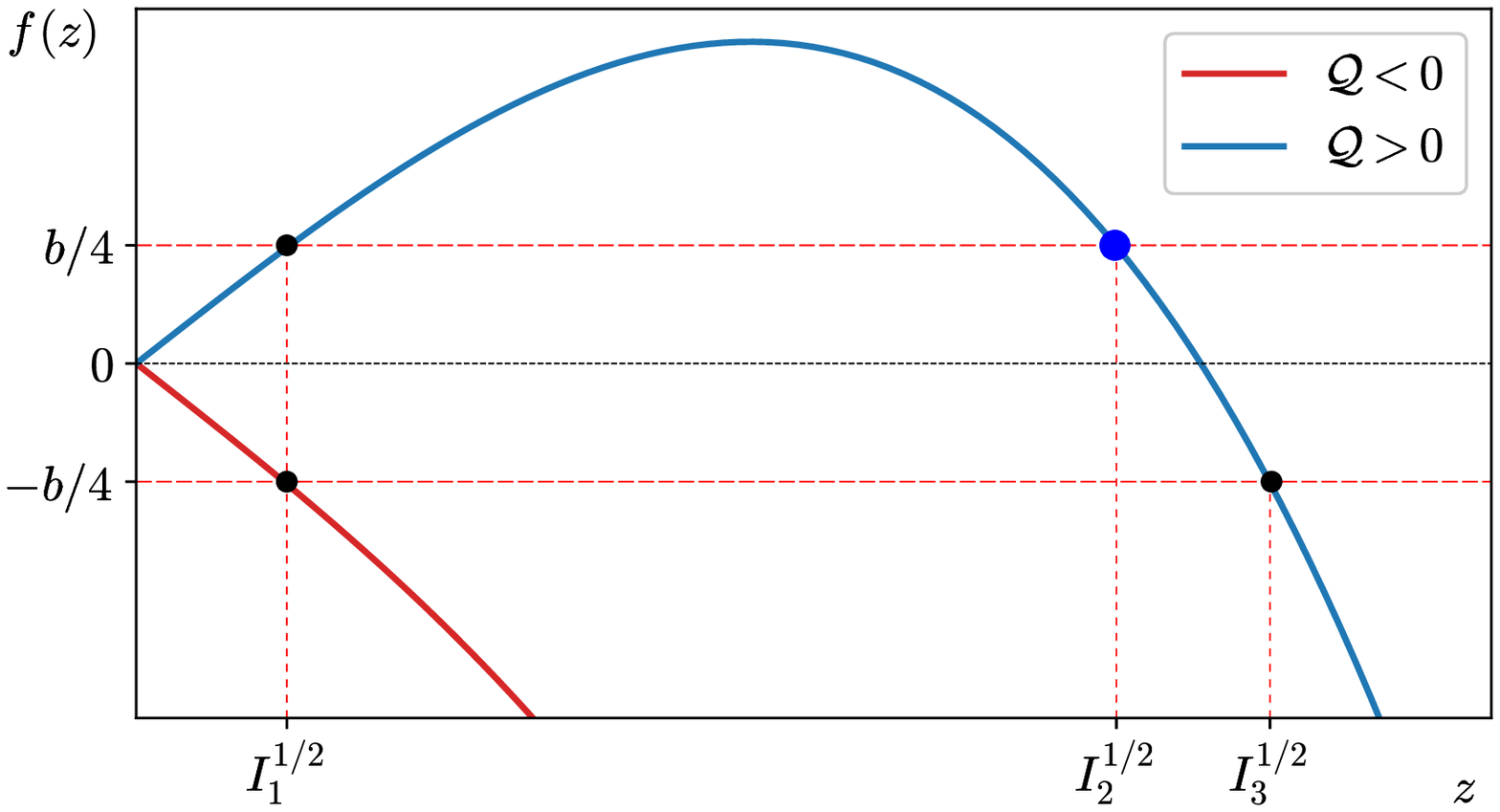} }
  \caption{Solutions of eqs. (\ref{eq:short}, \ref{eq:long}) for ${\cal P}>0$, $b>0$.}
  \label{fig:fz}
\end{figure}

To illustrate solutions of the last equations, we shall consider `abnormal' orbits, ${\cal P} > 0$. In case of ${\cal Q}>0$, function $f(z)$ have maximum $b_{\rm crit}/4$ (see Fig.\,\ref{fig:fz}), where
\be
  b_{\rm crit} \equiv \frac{8|{\cal Q}|}3 \left( \frac{\cal Q}{6\cal P} \right)^{1/2}\,.
  \label{eq:b_crit}
\ee
If $b<b_{\rm crit}$, eq. (\ref{eq:short}) has two solutions, otherwise there is no solution. Similarly, no solutions of this equation exist for ${\cal Q}<0$. One solution of eq. (\ref{eq:long}) corresponding to the closed orbit parallel to the long axis (L-orbit) exists for any signs of $Q$ and $(b-b_{\rm crit})$. The latter is often called the sequence $x_1$ (e.g., BT, sect. 3.3.2). The former solutions correspond to the closed orbits oriented parallel to the short axis of the potential (S-orbit): one with the lower eccentricity is stable (sequence $x_2$), and another one is unstable (sequence $x_3$).

\begin{figure}
\centering
  \centerline{\includegraphics[width=\linewidth, clip=]{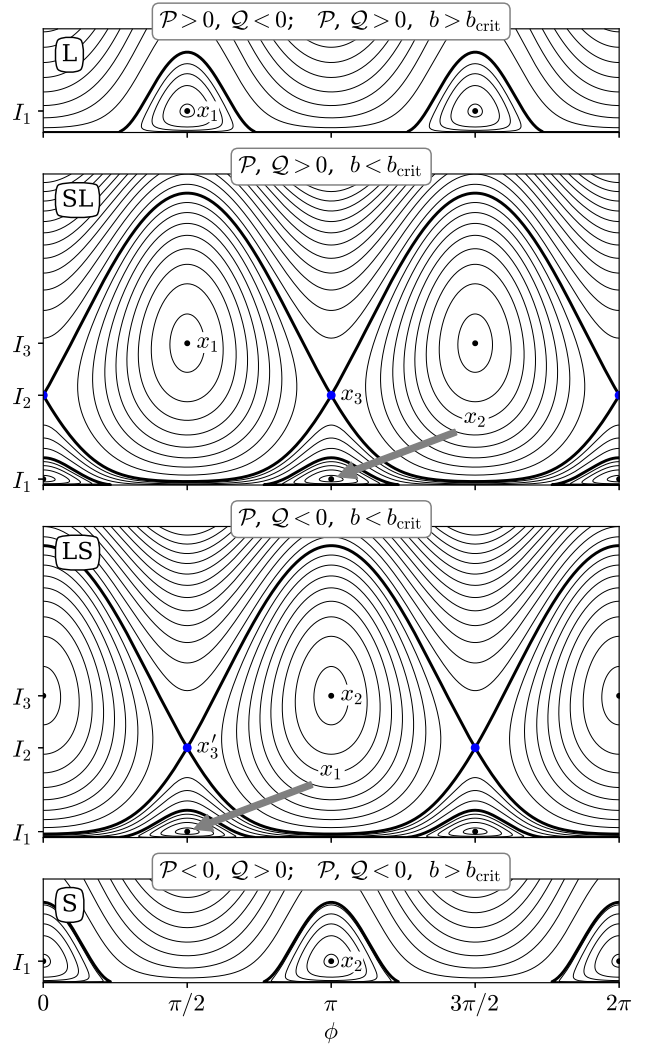} }
  \caption{Phase portraits in ($\phi$\,--\,$I$) planes of the averaged Hamiltonian (\ref{eq:cal_H_J}) for $b>0$. The closed L-orbits  correspond to $\phi=\pi/2, 3\pi/2$ (sequence $x_1$), the closed S-orbits -- to $\phi=0, \pi, 2\pi$ (sequence $x_2$). The saddle points $I_2$ (blue dots) correspond to unstable sequences $x_3$ and $x'_3$.}
  \label{fig:conts}
\end{figure}

All phase portraits are given in Fig.\,\ref{fig:conts}. The described above `abnormal' orbits, ${\cal P}>0$, give portraits (L) or (SL).\footnote{The first letter in the panel labelling shows the orientation of the closed stable orbit with smaller eccentricity.} Outside ILRs, only sequence $x_1$ is possible. Between ILRs the sequence $x_1$ becomes more eccentric and a new sequence of S-orbits $x_2$ may appear, if the bar amplitude is sufficiently small ($b<b_{\rm crit}$).
The phase portraits for the `normal' orbits, ${\cal P}<0$, are shifted by $\pi/2$ for the opposite sign of ${\cal Q}$: sequence $x_1$ is turned into $x_2$ (panels LS and S), low eccentric $x_2$ is turned into $x_1$, unstable sequence $x_3$ of S-orbits is turned into unstable sequence $x'_3$ of L-orbits (panel LS).
Changing of the sign of $b$ results only in the horizontal shift of all portraits by $\pi/2$.

\section{Examples}

\subsection{Power-law potentials}

\begin{figure}
\centering
  \centerline{\includegraphics[width=\linewidth]{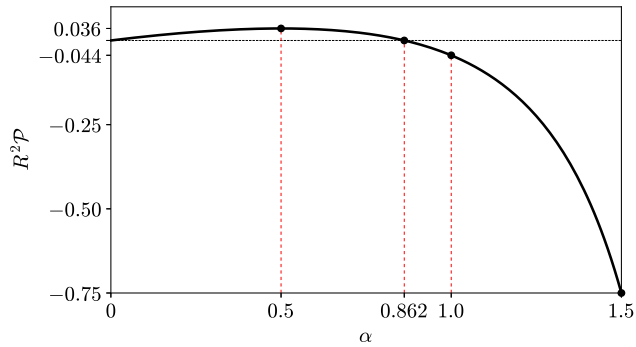} }
  \caption{Dependence of $R^2 {\cal P}$ versus $\alpha$ in the power-law potentials. }
  \label{fig:pa}
\end{figure}

This type of potentials include motion in Keplerian and harmonic potentials, and the Mestel disc with a flat rotation curve. Let's assume the angular speed in the form $\Omega(R) = \Theta R^{-\alpha}$. It is easy to show that
\be
  {\cal Q} =  \Theta\,R^{-\alpha}\cdot\Bigl(1-\sqrt{1-\alpha/2}\Bigr) - \Omega_{\rm p}\,,
\ee
and
\be
  {\cal P} =  \frac{\alpha}{2R^2} \left[ \left(\frac8{2-\alpha}\right)^{1/2} -\frac34 -\displaystyle\frac{\alpha}{6} - \frac2{2-\alpha} \right] \,.
  \label{eq:p}
\ee
Curve $R^2{\cal P}$ versus $\alpha$ is given in Fig.\,\ref{fig:pa}. It turns out that in power-law potentials, all nearly circular orbits could be either `normal' if $\alpha > 0.862$, or `abnormal' if $\alpha < 0.862$. Note that this boundary is close to $\alpha_{\rm BW} = 7/8$ of the \cite{1976ApJ...209..214B}  density profile $(\propto r^{-7/4})$.

The `normal' orbits naturally trap along the short axis of the potential inside ILRs ${\cal Q} > 0$ (portrait S), but they can be trapped along the long axis outside ILRs if $b<b_{\rm crit}$ (portrait LS). On the opposite, the `abnormal' orbits naturally trap along the long axis of the potential beyond the resonance (L) but can be trapped along the short axis if $b$ is sufficiently small (SL).

\subsection{The isochrone potential}

Consider the isochrone potential
\be
\Phi(r) = -\frac{GM}{a + (a^2+r^2)^{1/2}}\,,
\ee
for which the Jacobi integral reads:
\be
   H_0 = -2 G^2 M^2 t^{-2} - \Omega_{\rm p} L \,
\ee
where $t=2J_f+s$, $s= (L^2+4GMa)^{1/2}$. The LB-derivative of the precession rate can be obtained explicitely for any orbit~\citep[see also][]{1979MNRAS.187..101L}:
\be
  \frac{\p \Omega_{\rm pr}(L,J_f)}{\p L} = \frac{4G^2M^2}{s^3t^4} \left(4GMat - 3L^2 s \right) \,.
  \label{eq:piso}
\ee
In the limit of circular orbits (small $I$), one can use (\ref{eq:calP}) or put $t=L+s$ in eq. (\ref{eq:piso}).

\begin{figure}
\centering
  \centerline{\includegraphics[width=\linewidth, clip=]{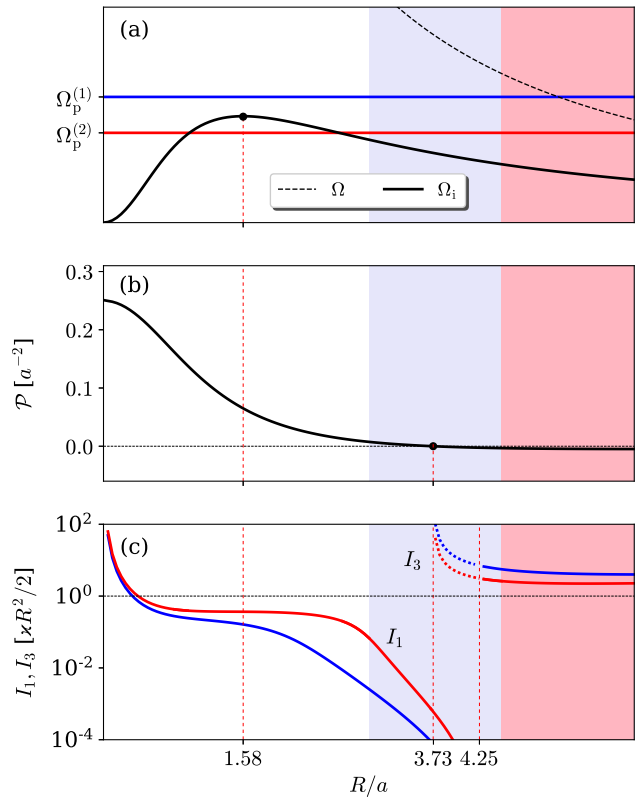} }
  \caption{Isochrone potential: (a) angular speed $\Omega$, $\Omega_{\rm i} \equiv \Omega - \varkappa/2$, and two pattern speeds $\Omega_{\rm p}^{(1)}$ and $\Omega_{\rm p}^{(2)}$; (b) LB-derivative (\ref{eq:calP}); (c) stationary points $I_{1}$, $I_{3}$ for the pattern speeds in (a) in units of $\varkappa R^2/2$ for model bar potential (\ref{eq:bar}), $\varepsilon=0.1$ (same colour coding). Solid/dotted lines in (c) show sequences $x_1$/$x_2$. Blue/pink shades show where the slowness assumption breaks down for $\Omega_{\rm p}^{(1)}$/$\Omega_{\rm p}^{(2)}$. Ticks at $1.58$, $3.73$ and $4.25$ mark maximum of $\Omega_{\rm i}$ and zeros of ${\cal P}$ and $b$, correspondingly. }
  \label{fig:iso}
\end{figure}

Fig.\,\ref{fig:iso}\,a shows angular speed $\Omega$, $\Omega_{\rm i} \equiv \Omega - \varkappa/2$, and two bar pattern speeds above and below the maximum of $\Omega_{\rm i}$. The slowness assumption is valid in the unshaded area for the larger pattern speed and breaks down further out. Similarly, for the smaller pattern speed, it breaks down in the pink area. Intersections of the pattern speed horizontal lines with angular speed $\Omega(r)$ give positions of corotation resonances, where the assumption is invalid.

The middle panel presents the LB-derivative ${\cal P}$ in units $a^{-2}$. It is positive inside $R = 3.73\,a$, and negative but vanishingly small outside this circle. This behaviour is natural and expected because of the damping factor $R^{-2}$ at large distances, see eq. (\ref{eq:p}).

Let's assume a model bar potential in the form:
\be
A(r) = -\varepsilon \, \frac{GM}{2a^2} \, r \, {\rm e}^{-r/a}\,.
\label{eq:bar}
\ee
From (\ref{eq:b}) we infer that sign of $b$ is determined by sign of expression $(1-R/a + 4\Omega/\varkappa)$,
which switches from positive to negative at $R = 4.25\,a$.

Panel (c) of Fig.\,\ref{fig:iso} illustrates the characteristic curves of sequences $x_1$ and $x_2$ for a matured bar only, $\varepsilon = 0.1$ (the bar amplitude is still small compared to the axisymmetric background). The stationary points $I_1$ and $I_3$ are obtained from (\ref{eq:short}) and (\ref{eq:long}). Curves $I_1$ are similar for these pattern speeds: despite two ILRs present for the red curve, the sequence $x_1$ corresponding to $I_1$ does not change to $x_2$, as it happens in the theory of weak bars. The family of S-orbits does not appear for $Q>0$, because $b$ exceeds the critical value $b_{\rm crit}$.

Solutions $I_3$ formally exist beyond $R=3.73$ but they obviously violate the epicyclic approximation. Note that it also breaks down for $I_1$ in the very centre, because $\varkappa R^2/2$ vanishes there.

\subsection{The Milky Way model}

The model we use here was elaborated in detail in our previous paper~\citep{PBJ16}. It consists of three components: thin exponential disc, S{\'e}rcic bulge and NFW halo. The disc is characterised by the radial scale $R_{\rm d} =2.9$ \,kpc, vertical scale $z_{\rm d}=300$\,pc and mass $M_{\rm d} = 4.2\cdot 10^{10}\,{\cal M}_\odot$ (solar mass). The bulge has a weak cuspy density profile in the centre $\rho_{\rm b} \propto r^{-1/2}$, and mass $M_{\rm b} \approx 10^{10}\,{\cal M}_\odot$. The total circular velocity is bulge-dominated at radii $R \lesssim 2.5$\,kpc, and halo-dominated at $R > 9$\,kpc. At radius $R = 6$\,kpc, where the disc contribution peaks, the force from the halo is about $2/3$ of the force from the disc in the galactic plane.

N-body simulations show bar instability producing a bar rotating with pattern speed $\Omega_{\rm p} = 55$\,km/s/kpc. A bar amplitude grows exponentially in time with a small growth rate $\gamma \sim 0.07 \,\Omega_{\rm p}$
and saturates at the level 10\,...\,20 per cent of the axisymmetric background. After that, the amplitude stays nearly constant, but the bar pattern speed gradually decreases.

It is well known that ILR damps spiral waves~\citep{Mark74}\footnote{In the purely linear theory. The nonlinear effects near ILR can deactivate damping~\citep[][]{2019MNRAS.483..692P}.}. Through this effect, the bar formation is suspended in flat disc galaxies. However, the bar can still be formed if ILR radius is comparable with or smaller than the disc vertical scale. Moreover, the bar pattern speed and the growth rate can be reproduced well from the linear perturbation theory for flat discs, if one uses an angular speed ${\overline\Omega}$ averaged over vertical axis $z$, instead of in-plane $\Omega$ calculated from the total axisymmetric potential~\citep{PBJ16}.

\begin{figure}
\centering
  \centerline{\includegraphics[width=\linewidth, clip=]{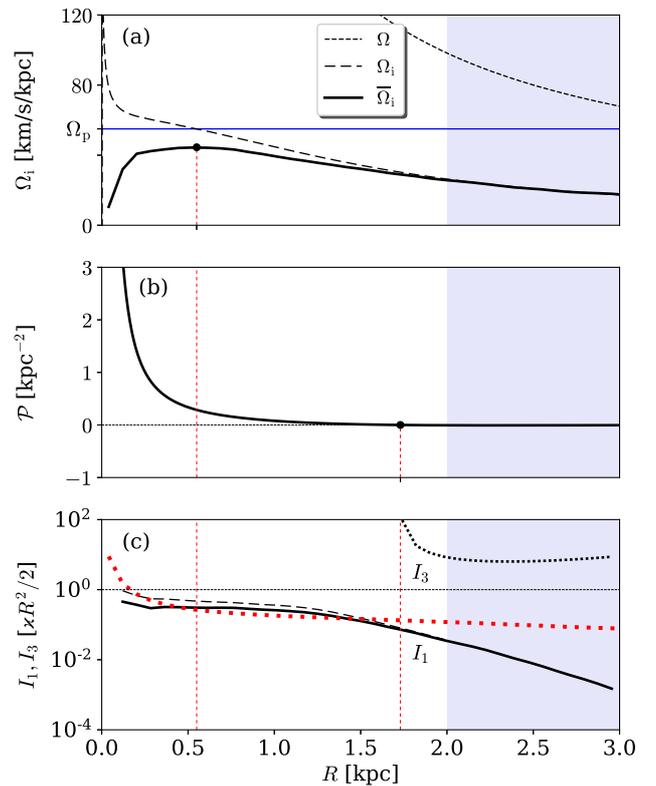} }
  \caption{Same as Fig.\,\ref{fig:iso}\, for the Milky Way model: (a) $\Omega_{\rm i}$ (long dashes) now marks the in-plane quantity, while  ${\overline\Omega}_{\rm i}$ (solid) marks the quantity averaged over the vertical axis (see text); pattern speed of the bar $\Omega_{\rm p}$, as obtained in N-body simulations~\citep{PBJ16}; (c) stationary points $I_{1}$, $I_{3}$ for the maximum bar potential (the line type coding corresponds to panel (a) and Fig.\,\ref{fig:iso}\,c). Red dots show $\sigma^2/\varkappa^2 R^2$ -- typical normalised radial actions populated in the disc ($\sigma$ is the radial velocity dispersion). Ticks at $0.55$ and $1.73$ mark maximum of ${\overline\Omega}_{\rm i}$ and zero of ${\cal P}$. }
  \label{fig:mw}
\end{figure}

Fig.\,\ref{fig:mw}\,a shows the in-plane $\Omega$ and $\Omega_{\rm i}$, $z$-averaged ${\overline\Omega}_{\rm i}$, and the initial bar pattern speed $\Omega_{\rm p}$. A vertical dashed line at $R = 0.55$\,kpc marks the maximum of  ${\overline\Omega}_{\rm i}$. Curve ${\cal P}(R)$ on panel (b) is calculated using ${\overline\Omega}$. Similar to Fig.\,\ref{fig:iso}\,b, it is positive in the centre, and is slightly negative beyond $R=1.73$\,kpc.

Panel (c) presents characteristic curves of sequences $x_1$ and $x_2$ for ${\overline\Omega}_{\rm i}$ and the maximum bar amplitude $b$ obtained from N-body snapshots (in particular, $T=1.3$\,Gyr). These curves are qualitatively similar to those shown on Fig.\,\ref{fig:iso}\,c. In particular, the central part is populated with $x_1$ orbits only. The sequence $x_2$ formally obtained beyond 1.73 kpc consists of too eccentric orbits to be represented in the galactic disc (red dots show typical radial actions populated in the disc). Notably, the curves of sequences for the in-plane $\Omega$ come almost the same as for ${\overline\Omega}_{\rm i}$ (dashed line for $I_1$, not shown for $I_3$).

\section{Discussion and summary}
\label{sec:summary}

Using the standard technique of finding stationary points of the Hamiltonian, we show that orientation of orbits is governed by the signs of the precession rate ${\cal Q}$, of the LB-derivative\footnote{The derivative over angular momentum at constant adiabatic invariant $J_f$.} of the precession rate ${\cal P}$, and of the orbital responsiveness to the bar potential $b$. This is in accordance with our previous works based on the matrix methods of the linear perturbation theory that show the importance of the sign of the precession rate for radial-orbit \citep[][]{2010AstL...36...86P, 2015AstL...41....1P, 2015MNRAS.451..601P, 2017MNRAS.470.2190P}
and loss cone instabilities \citep[][]{2007MNRAS.379..573P, 2008MNRAS.386.1966P,2010AstL...36..175P}.

These new results extend the theory of bar formation by~\citet{1979MNRAS.187..101L} that classifies all disc orbits using only one of these three parameters -- the sign of ${\cal P}$. A majority of orbits consist of so-called `normal' orbits characterised by the negative sign of ${\cal P}$. It tends to align in the direction perpendicular to the bar. The smaller fraction of orbits populating the central part of the disc called `abnormal', for which ${\cal P}>0$, aligns with the bar thereby reinforcing it.

\begin{figure}
\centering
  \centerline{\includegraphics[width=\linewidth]{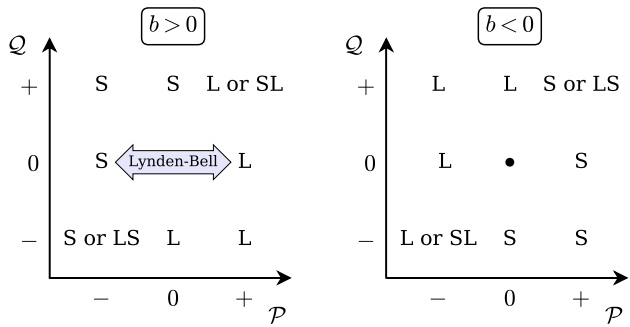} }
  \caption{The phase portraits for $b>0$ (left) and $b<0$ (right). Double arrow marks two cases suggested by~\citet{1979MNRAS.187..101L}.}
  \label{fig:diag}
\end{figure}

The progress of the current work is two-fold. First, the inclusion of the second parameter ${\cal Q}$ leads to a variety of combinations for orbits to align with the bar. Fig.\,\ref{fig:diag} condenses the phase portraits types for any possible combination. The portraits given in Fig.\,\ref{fig:conts} contain the stationary points corresponding to the well-known sequence $x_1$ of the orbits aligned parallel to the long axis, as well as stable sequence $x_2$ and unstable sequence $x_3$ of the orbits aligned parallel to the short axis of the potential perturbation.

Second, analysis of the realistic models shows that although formally ${\cal P}$ is negative outside the central region, it is small in the absolute value. This discriminates the role of so-called `normal' orbits to destroy the bar. Indeed, the presence of ${\cal Q}$-term allows to put ${\cal P}=0$ in this region, thus the portraits for ${\cal P} <0$ can be essentially ignored.

Section 3.3 presents a simple Milky Way model with a weak cusp in the centre $\rho \propto r^{-1/2}$ and nearly flat rotation curve outside $R = 1.5\, ...\, 2$\,kpc circle~\citep[see Fig.\,7 of][]{PBJ16}. In the inner disc, the positive sign of ${\cal P}$ plays a major role in determining the orientation of orbits along the potential well. However, if ${\cal Q} > 0$, our theory predicts a family of short-axis orbits (S-orbits) for small $b$ (portrait SL). This presumably explains the well-known phenomenon \citep[e.g.,][]{1993A&A...271..391C} that bars in N-body simulations have pattern speeds larger than the maximum of $\Omega_{\rm i}$  (i.e. ${\cal Q}<0$) because S-orbits in case of ${\cal Q} > 0$ immediately destroy low amplitude bar-like perturbations. Only perturbations with ${\cal Q}<0$ can be reinforced by trapping the orbits along the potential well. Remarkably, the matured bar can sustain the pattern speed decrease below the maximum of $\Omega_{\rm i}$, because for large $b$ only L-orbits are possible.

In the outer disc beyond point ${\cal P}=0$, orbits continue to add to the bar unless the bar pattern speed is too low so that the orbits find themselves between two ILR's, i.e. ${\cal Q}>0$.

In the theory of weak bars \citep{1976ApJ...209...53S, 1993RPPh...56..173S}, the epicyclic approximation (\ref{eq:epic}) is used to derive orientations of nearly circular orbits. Below we follow sect. 3.3.3 of BT to compare their closed loop orbits with ours. Their `epicyclic radius' is
\be
  C_2 = -\frac{A}{R \Delta} \left(\frac{R}{A}\frac{\d A}{\d R} + \frac{2\Omega}{\Omega-\Omega_{\rm p}} \right) \,,
\ee
where $\Delta = \varkappa^2 - 4\,(\Omega-\Omega_{\rm p})^2$. Near the resonance $ \varkappa \approx  2\,(\Omega-\Omega_{\rm p})$, so $\Delta$ can be substituted by $-4{\cal Q}\varkappa$. The corresponding radial action is then
\be
  I = \frac{\varkappa}2 C_2^2 \approx \frac{\varkappa}2 \frac{A^2}{16 R^2 Q^2 \varkappa^2} \left(\frac{R}{A}\frac{\d A}{\d R} + \frac{4\Omega}{\varkappa} \right)^2\,.
\ee
The last expression coincides with our stationary point $I_1$ for L-orbits obtained from (\ref{eq:long}) outside ILRs, provided ${\cal P} = 0$ \citep[see also][]{1981ApJ...243.1062G}.

The `epicyclic radius' $C_2$ formally changes its sign at the inner and outer ILRs due to $\Delta$, resulting in appearance of $x_2$ sequence of orbits perpendicular to the potential well ($x_1$-$x_2$-$x_1$ sequence in Fig.\,3.20 of BT). From our theory it follows (Fig.\,\ref{fig:mw}\,c) that orbits' orientation along the potential well is retained between the resonances for large bar amplitudes (c.f. Fig.\,3.18 of BT). Note that in case of the weak bar, the L-orbit family continues smoothly across the resonances, but additional S-orbit family appears around smaller $I_1$.

The parameter $b$ becomes negative at radius $R_b$ where the round bracket in (\ref{eq:b}) vanishes. The physical meaning of this radius is the last closed orbit of $x_1$ sequence, so it can be used as a clearly detectable proxy of the bar length~\citep[see also][]{2006ApJ...637..214M}. At $R_b$ the LB-derivative is likely to be nearly zero and the precession rate ${\cal Q}<0$.

Summarizing the above, two quantities specify the direction of orbit's trapping with respect to the potential well: the precession rate ${\cal Q}$ and the LB-derivative of the precession rate ${\cal P}$. Their interplay allows us to explain the features of bar formation observed in N-body simulations and reconcile the Lynden-Bell theory with the theory of weak bars. The third parameter $b$ describing the orbits' responsiveness to the potential, may alter the orbital alignment, but this may only happen well outside the central region.

\section*{Acknowledgments}
This work was supported by the Deutsche Forschungsgemeinschaft (DFG, German Research Foundation) -- Project-ID 138713538 -- SFB 881 (``The Milky Way System'', subproject A06), by the Volkswagen Foundation under the Trilateral Partnerships grant No. 97778, by RFBR grant 20-52-12009, Foundation for the advancement of theoretical physics and mathematics ``Basis'' and by Department of Physical Sciences of RAS, subprogram `Interstellar and intergalactic media: active and elongated objects'. The work also was partially performed with budgetary funding of Basic Research program II.16 (Ilia Shukhman).

\section*{Data Availability}

Data underlying this article will be shared on reasonable request to the authors via epolyach@inasan.ru. Data related to the initial conditions may be reproduced via the publicly available software {\texttt GalactICS}.

\bibliographystyle{mnras}


\end{document}